\documentclass[final]{svjour2}
\usepackage{graphicx}
\usepackage{rotating}
\usepackage{amssymb}
\usepackage{mathptmx}
\usepackage[numbers]{natbib}

\makeatletter
\journalname{Journal of Low Temperature Physics}

\bibpunct{}{}{,}{s}{}{,}

\begin{document}

\newcommand{\hdblarrow}{H\makebox[0.9ex][l]{$\downdownarrows$}-}
\title{Probing Bogoliubov quasiparticles in superfluid $^3$He with a `vibrating-wire like' MEMS device}

\author{M.~Defoort$^1$ \and S.~Dufresnes$^1$ \and S.L.~Ahlstrom$^2$  \and D.I.~Bradley$^2$ \and R.P.~Haley$^2$
\and A.M.~Gu\'enault$^2$ \and E.A.~Guise$^2$ \and G.R.~Pickett$^2$ \and M.~Poole$^2$ \and A.J.~Woods$^2$
\and V.~Tsepelin$^2$ \and S.N.~Fisher$^2$$^\dagger$ \and H.~Godfrin$^1$ \and E.~Collin$^1$}
\institute{$^1$ Universit\'e Grenoble Alpes, Institut NEEL, F-38000 Grenoble, France\\
\email{eddy.collin@grenoble.cnrs.fr} \\
$^2$ Department of Physics, Lancaster University, Lancaster, LA1 4YB, UK\\
\email{v.tsepelin@lancaster.ac.uk} \\
$\dagger$ Deceased 4 January 2015}

\date{21.06.2015}
\maketitle

\begin{abstract}

We have measured the interaction between superfluid $^3$He-B and a micro-machined goalpost-shaped device at temperatures below $0.2\,T_c$.
The measured damping follows well the theory developed for vibrating wires, in which the Andreev reflection of quasiparticles in the flow field around the moving structure leads to a nonlinear frictional force. At low velocities the damping force is proportional to velocity while it tends to saturate for larger excitations. Above a velocity of 2.6\,mms$^{-1}$ the damping abruptly increases, which is interpreted in terms of Cooper-pair breaking. Interestingly, this critical velocity is significantly lower than reported with other mechanical probes immersed in superfluid $^3$He. Furthermore, we report on a nonlinear resonance shape for large motion amplitudes that we interpret as an inertial effect due to quasiparticle friction, but other mechanisms could possibly be invoked as well.

\end{abstract}

\section{Introduction}

In the sixties, Tough, McCormick and Dash showed that taut vibrating wires are excellent probes for fluids having low viscosities\cite{taught1,taught2}, for example He-II. The devices were driven by a small a.c. current in a d.c. magnetic field (Laplace force), and the motion detected through the induced voltage (Lenz's law). Later on in the eighties, this technique was applied to superfluid $^3$He, using very thin superconducting wires shaped as semi-circular resonators\cite{hall,pickett}. At extremely low temperatures ($T \ll T_c$), the effective viscosity of the superfluid that corresponds to the quasiparticle density has been measured by these mechanical probes. Many experiments since then have been using this technique, for example in the development of black-body radiators\cite{shaunBB} and their application for particle detection\cite{darkmatter1995,ultima}.
The main limiting factor has been found to be the intrinsic damping of the probes.

Since 2004 many groups have been also using the very practical quartz tuning forks to probe helium properties\cite{buu,Blau}. These piezoelectric devices are extremely cheap, easy to handle and do not require a magnetic field to operate. They have proven to be as efficient as standard vibrating wires to measure extremely low  quasiparticle densities, or equivalently temperatures\cite{roch}.
However, in the viscous limit the flow between the two prongs oscillating in anti-phase is far more complex, and much less easy to handle compared to the flow around a single wire. The dynamics relies on the piezoelectric effect and the piezoelectric constant needs to be calibrated\cite{TFJLTP2010,PRBVik}.
For commercial devices, the geometry is fixed and material quality needs to be validated by a careful selection of the probes.

\begin{figure}
\begin{center}
\includegraphics[width=0.9\linewidth,keepaspectratio]{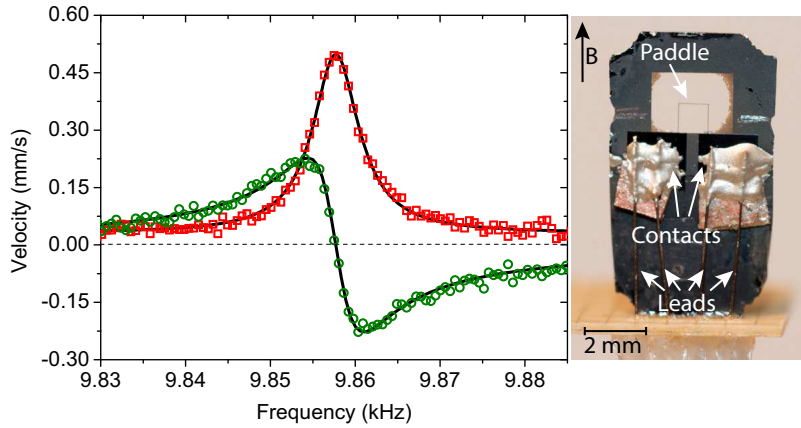}
\end{center}
\caption{(Color online) Left: first flexure resonance in superfluid $^3$He-B around 155\,$\mu$K. Red squares and green circles correspond to the in-phase ($X$) and out-of-phase ($Y$) signal, respectively. Lines are Lorentzian fits with a resonance frequency $f_0$ of 9.857\,kHz and a linewidth $\Delta \! f$ of 6.6\,Hz. Right: an image of a goalpost silicon device mounted in a $^3$He experiment.}
\label{fig1}
\end{figure}

Research and technologies associated with Micro-Electro-Mechanical Systems (MEMS) and micro-fluidics have been continuously expanding over the last de\-ca\-des\cite{memsbook,microfluidbook}. Monocrystalline silicon has proven to be an extremely good mechanical material, enabling mechanical resonances with $Q$s of the order of a million at low temperatures \cite{Qsilicon}. Oscillating square-type cantilevers have been used for viscosity measurements with great success in classical fluids \cite{bullard}, and theoretical developments have been done to perfectly characterize both their mechanical modes \cite{eddymodes} and the flow fields they generate \cite{saderflow}. Furthermore, using basic lithography techniques, metal-coated silicon-based devices offer a unique versatility in both shapes and sizes, down to the nano-mechanical (NEMS) scale \cite{nems}.

For this reason, the Grenoble group proposed to engineer 'vibrating-wire like' MEMS structures \cite{physicaB}. These devices can be rather easily scaled down to NEMS, with applications in quantum fluids physics research \cite{eddyQFSJLTP}. Their mechanical properties are extremely good, with high quality factors and the possibility to use many modes of the structure \cite{eddymodes}.
In the present paper, we report on the first measurements obtained with such a MEMS device immersed in superfluid $^3$He-B.
\begin{figure}
\begin{center}
\includegraphics[width=0.7\linewidth,keepaspectratio]{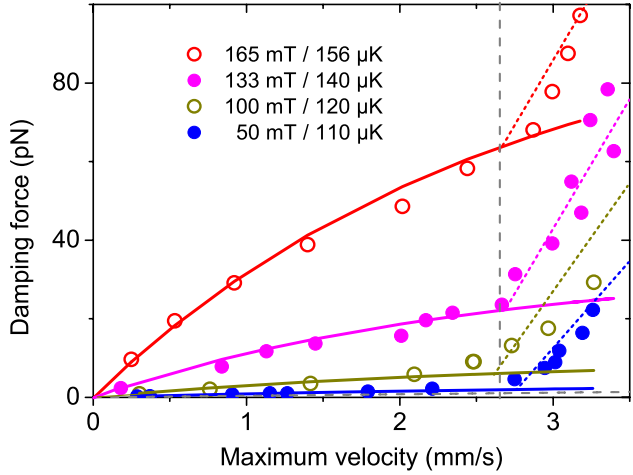}
\end{center}
\caption{(Color online) Damping force $F_d$ versus maximum velocity $v_m$ at various temperatures in $^3$He-B (see legend). The vertical dashed line at $\approx$2.6\,mm/s marks the measured critical velocity. Below this velocity the solid lines are fits to the Andreev reflection model, while above the critical velocity, in the pair-breaking regime, the dotted lines are guides for the eyes. For the details see text.}
\label{fig2}
\end{figure}

\section{Experimental results}

The device we use is a goalpost-shaped MEMS\cite{physicaB}, with feet and paddle about 1\,mm long, thickness and width about 20\,$\mu$m. It is made of monocrystalline silicon, covered by a thin (200\,nm) layer of aluminum. The sample has been mounted in a Lancaster cell \cite{lancatsercell}. It is held in place using Stycast 1266 glue, and electrical contacts onto the chip have been made using Epotek silver epoxy. The Fig.~\ref{fig1} shows an image of our device. The device has been cooled down to 0.15\,$T_c$ in superfluid $^3$He at 0\,bar, using the final magnetic fields of the nuclear demagnetization protocol\cite{pickett}.
In these conditions, the stable superfluid phase is $^3$He-B.

Our MEMS wire is driven using a magnetomotive scheme as a typical vibrating wire\cite{eddyJLTP}. A magnetic field $B_0$ (typ. $< 200~$\,mT) is applied perpendicularly to the paddle and parallel to the feet of the MEMS structure (see Fig.~\ref{fig1}). When the frequency $\omega$ of a.c. current $I_0 \cos (\omega \, t)$ flowing through the MEMS via the metallic layer from the electric contacts is close to the first flexural mode resonance, the MEMS structure oscillates due to the Laplace force $I_0 l B_0$, where $l$ is the length of the paddle. The motion is detected through the induced voltage $l B_0 v$, where $v$ is the velocity of the paddle bar. A typical resonance line measured using a lock-in amplifier is shown in Fig.~\ref{fig1} together with a Lorentzian fit: we thus extract the height of the peak $H$, its full-width-at-half-height $\Delta \! f$ and the resonance frequency $f_0$. We easily verify that $H \propto 1/\Delta \! f$.

For low velocities, the thermal damping on the moving object is governed by the Andreev reflection of quasiparticles and is nearly constant \cite{pickett,shaun,andreev}, while at larger velocities, the damping force tends to saturate until the pair-breaking velocity is reached\cite{shaun}. The onset of pair-breaking (dashed vertical in Fig.~\ref{fig2}) observed at $\approx$2.6\,mm/s for the MEMS device is rather low in comparison with vibrating wires and quartz tuning forks at 0\,bar that exhibit a critical velocity of $\approx$9\,mm/s. We attribute such low value to the flow enhancement around of the sharp square edges of the MEMS probe, generated by the Reactive Ion Etching (RIE) fabrication \cite{eddyJLTP}.

\begin{figure}
\begin{center}
\includegraphics[width=0.85\linewidth,keepaspectratio]{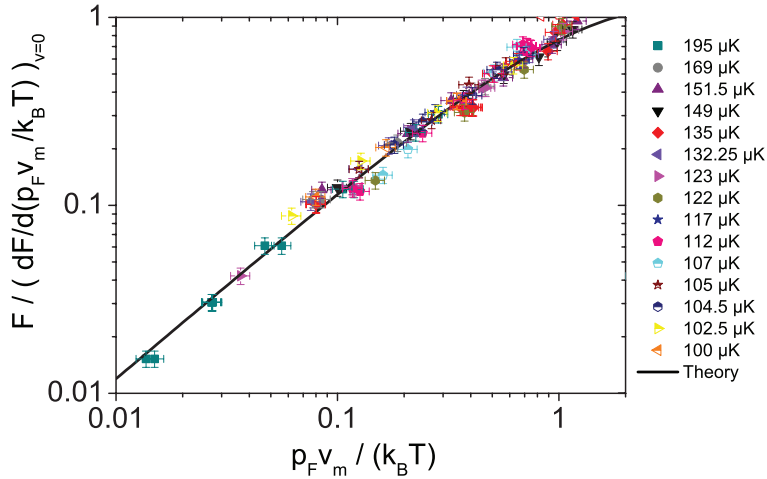}
\end{center}
\caption{(Color online) Scaled damping force versus scaled velocity for all measured temperatures. The black line is a fit to Eq.~(\ref{eq1}) leading to the definition of $\lambda$ and $\gamma$. Pair-breaking has been omitted in this plot.}
\label{fig3}
\end{figure}

The saturation of the thermal damping force $F_d$ due to the Andreev reflection mechanism at higher velocities of the moving object can be described as:
\begin{equation}
F_d \propto \gamma \left[1- \exp \left( -\lambda \frac{p_F\,v_m}{k_B\,T} \right)\right],
\label{eq1}
\end{equation}
where $p_F$ is the Fermi momentum, $v_m$ is the maximum velocity of the paddle structure, $k_B$ is the Boltzmann's constant and $T$ is the temperature. Two geometrical constants $\gamma$ and $\lambda$ are determined using a fit to the data shown in Figs. \ref{fig2} and \ref{fig3}. We find $\lambda \approx 1$ which is consistent with vibrating wire results \cite{shaun}. The constant $\gamma$ that defines the actual quasiparticle cross section of the device is found to be $\approx 0.7$ by scaling the measured thermal contribution to the linewidth of the MEMS against the one of standard vibrating wires \cite{roch}. Contrasting $\gamma \approx 0.7$ to values of 0.3 for vibrating wires \cite{shaunBolo} and 0.6 for quartz tuning forks \cite{roch}, shows that the flat, single-bar geometry made from monocrystalline silicon is more sensitive to the fluid's thermal excitations than other devices.

In Fig. \ref{fig3} we demonstrate a good agreement between measured data and the fit to Eq.~(\ref{eq1}) (black line) over the full range of explored velocities and temperatures. The error bars in Fig. \ref{fig3} are about $\pm10\,\%$ and reflect our measurement resolution on height $H$ or width $\Delta \! f$ in the present experimental run. The observed finite resolution is the result of a nonlinear behaviour of the device, a non-optimal cryogenic amplification stage and a small series ohmic resistance, caused by unexpectedly low-quality silver epoxy joints.

\begin{figure}
\begin{center}
\includegraphics[width=0.7 \linewidth,keepaspectratio]{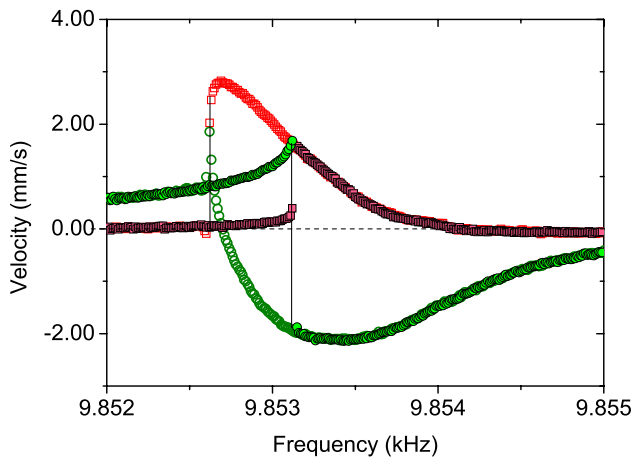}
\end{center}
\caption{(Color online) Measured resonance line at about 100\,$\mu$K, with a maximum velocity of about 2.6\,mm/s. The lineshape is strongly nonlinear with bistability (between the two vertical lines). The color code is the same as in Fig.~\ref{fig1} (for empty squares/circles), with filled (empty) symbols corresponding to upwards (downwards) frequency-sweeps.}
\label{fig4}
\end{figure}

Surprisingly, in $^3$He the resonance lineshape was nonlinear even for rather low velocities (corresponding to small maximum displacements) (see Fig. \ref{fig4}).
Nonlinearity makes the measurements tedious: complete frequency sweeps have to be measured for each excitation in order to prevent bistable switchings and truncated datasets. Furthermore, because of the smallness of the displacement amplitudes, the detected signal is also rather small. All of the above facts lead to the error bars shown in Figs. \ref{fig3} and \ref{fig5}.

\begin{figure}
\begin{center}
\includegraphics[width=0.7 \linewidth,keepaspectratio]{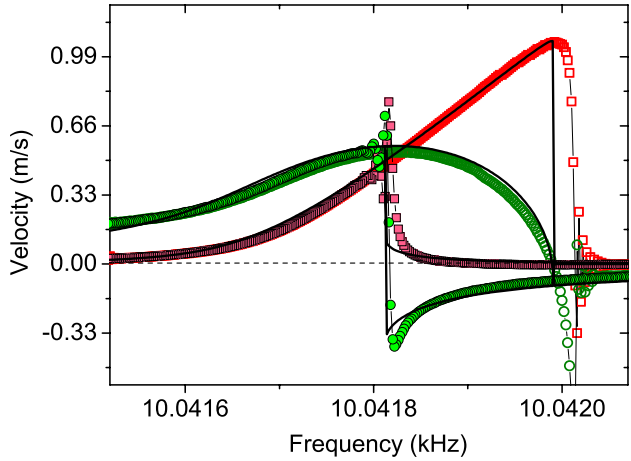}
\end{center}
\caption{(Color online) Resonance line measured in vacuum at 4.2$~$K. The maximum velocity was about 1.0$~$m/s. The color code is the same as in Fig. \ref{fig4}, and the black lines are a fit to the standard Duffing model \cite{eddyPRBnlin}. The visible difference between fit and data is due to the finite time constant of the lock-in detector. The fit leads to the linear parameters (frequency $f_0=10\,041.77$\,Hz and linewidth $\Delta \!f = 45$\,mHz), and to the Duffing coefficient $\beta = +0.45\times10^9$\,Hz/m$^2$ (consistent with literature).}
\label{fig4bis}
\end{figure}

When the lineshape is nonlinear, the dissipation cannot be inferred from a line\-width measurement. However, the simple relation between height $H$ and line\-width parameter $\Delta \! f$ still holds, and it is thus possible to infer the linear dissipation coefficient (in units of Hertz) by measuring the maximum height of the resonance peak \cite{eddyPRBnlin}. But no convincing full fit of the raw data, Fig.~\ref{fig4}, could be produced because the source of the nonlinearity is not yet clearly identified. We can rule out device-dependent nonlinear effects: the intrinsic dominant non-linearity is of geometrical origin and is thus temperature-independent\cite{eddyJLTP,eddyPRBnlin}. A resonance line measured in vacuum at 4.2\,K is shown in Fig.~\ref{fig4bis} for a large motion amplitude. The non-linearity is much smaller and is positive (frequency hardening), compared to the non-linearity measured in $^3$He-B. Fig.~\ref{fig5} shows the position of the resonance peak as a function of maximum velocity to highlight the nonlinear effect in the superfluid.

We can only speculate about the source of the nonlinearity. It could be the signature of the inertial nonlinear component which should accompany the nonlinear damping of quasiparticles, or quasiparticle emission or turbulence nucleation. One would need a quantitative estimate of these effects to compare with Fig.~\ref{fig5}, and explain its origin. At present, the simplest fit to Fig.~\ref{fig5} that can be produced is obtained via a nonlinear Duffing-like inertial term of the type $m_0(1+m_2 \, v_m^2/(2 \pi f_0)^2)$ replacing the mode mass $m_0$. The nonlinear coefficient $m_2$, which would be related to superfluid properties, is then linked to the fit Duffing coefficient $\beta$ in Fig.~\ref{fig5} (full line) by the relation: $\beta=-\frac{1}{4} f_0 \, m_2$. We extract for the nonlinear mass $m_0 m_2/(2 \pi f_0)^2 = 4.1\times10^{-8}$\,kg/(ms$^{-1}$)$^2$. Interestingly, rather different frequency-softening nonlinear features were reported at ultra-low temperatures on a conventional vibrating wire, and at the time were interpreted as the interplay between the wire and the $^3$He-B texture\cite{PRL83}.

\begin{figure}
\begin{center}
\includegraphics[width=0.7\linewidth,keepaspectratio]{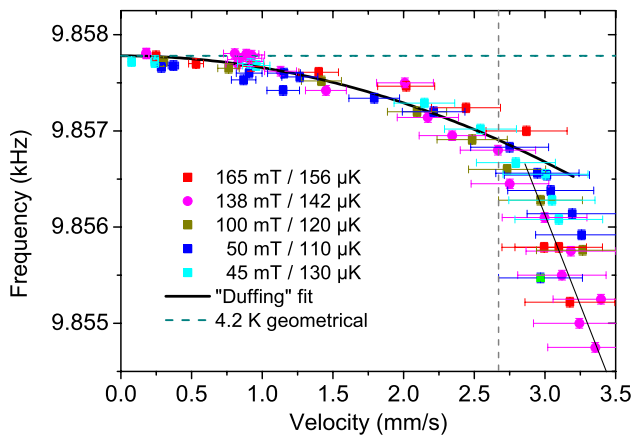}
\end{center}
\caption{(Color online) Position of the resonance peak as a function of maximum velocity at different temperatures in $^3$He-B (see legend). The dashed vertical marks the pair-breaking velocity. Below pair-breaking, the fit is a simple quadratic Duffing law with $\beta = -4.7\times10^{14}$\,Hz/m$^2$. Above the pair-breaking limit, the line is a guide to the eyes. The horizontal dashed line represents the geometrical (and positive) quadratic Duffing fit from Fig.~\ref{fig4bis}.}
\label{fig5}
\end{figure}

\section{Conclusions}

We have measured in superfluid $^3$He at ultra-low temperatures the damping due to Andreev reflection of quasiparticles on a MEMS goalpost-shaped vibrating-wire like structure. We demonstrate that the nonlinear damping follows the same model as for vibrating wires, with the same geometrical parameter $\lambda\approx1$ and a larger $\gamma\approx0.7$. We thus show that this probe is more sensitive to thermal excitations than other structures immersed in superfluid $^3$He. Mentioning the versatility of microfabrication technologies and the very high vacuum mechanical $Q$ factors reachable for these structures (here, above 0.2 million), these new probes are to date the best technology for the lowest temperature measurements and highest resolution in bolometric particle detection in superfluid $^3$He.
However, the critical velocity is much lower than for other conventional structures and a nonlinear behaviour is observed at ultra-low temperature, which we interpret as a nonlinear inertial effect. Further experiments are clearly needed to contrast the onset of extra damping at critical velocity and to understand the nature of non-linearities reported with MEMS and vibrating wires. In particular, measurements at different fluid densities (pressures) should be performed to see how the observed properties scale with superfluid parameters.
Finally, a natural development of this research for the future is to scale down these structures toward NEMS and see how the superfluid properties are modified when probed on a scale of order $\xi_0$.

\begin{acknowledgements}
We would like to thank B.~Fernandez, T.~Fournier, C.~Blanc and O.~Bourgeois for help with the fabrication of the devices. We acknowledge the support from MICRO\-KELVIN, the EU FP7 low temperature infrastructure grant 228464, the 2010 ANR French grant QNM n$^\circ$ 0404 01, and the UK EPSRC grant EP/I028285/1.
\end{acknowledgements}

\end{document}